# Dimensional Complexity & Algorithmic Efficiency

**Alexander Odilon Ngu**

Ngu Energy, Boston, USA
Email: amayangu@bu.edu;
nguenergy@gmail.com

## Abstract

This paper uses the concept of algorithmic efficiency to present a unified theory of intelligence. Intelligence is defined informally, formally, and computationally. We introduce the concept of Dimensional complexity in algorithmic efficiency and deduce that an optimally efficient algorithm has zero Time complexity, zero Space complexity, and an infinite Dimensional complexity. This algorithm is used to generate the number line.

## 1. Introduction

In 1931, Kurt Gödel published his first incompleteness theorem and changed logic forever [1]. Gödel's first incompleteness theorem stated that no single formal system could completely define all mathematical truths and that physical theories depend on mathematical theories, which are incomplete because they contain unprovable statements. Thus, all physical theories are incomplete [2]. In his 1936 paper "An unsolvable problem of elementary number theory", Alazo Church formalized the first theory of a computable function and an incomputable function [3]. In 1936, in his paper "On computable numbers, with an application to the Entscheidungsproblem", Alan Turing formalized an algorithmically computable and incomputable function [4]. In 1947, Russian mathematician A.A. Markov refined the definition of the algorithm with his concept of a normal algorithm [5]. A normal algorithm is applicable to alphabetic notations, which in mathematics are defined as a finite set of distinguishable notations or symbols [5]. Normal algorithms are akin to Turing machines or partial recursive functions [6].

Gödel's incompleteness theorems, Turing's incomputability theorem, and Markov's normal algorithm were monumental because they pointed out an encapsulating boundary around mathematical theories, finitary algorithms, and computable functions resulting from incompleteness and inconsistency and or incomputability. This paper aims to show that algorithmic efficiency can be more completely determined by not only accounting for the complexity of the computation that an algorithm carries out but also the complexity of the algorithm itself. This understanding is used to define intelligence as an abstraction or formalization of generality.

## 2. Complexity

How do we compare algorithms to measure their efficiency? Typically, algorithmic efficiency is measured by complexity. Computational complexity can be measured in space and time. *Space complexity* denotes space required for execution, and *Time complexity* denotes the number of operations required to complete execution [7]. *Time complexity* is measured by the number of iterations it takes for an algorithm to execute, and *Space complexity* is measured by the amount of memory or space required for the algorithm to execute.

Theoretically, minimizing *Time complexity* and *Space complexity* should increase the efficiency of an algorithm. With this understanding, we can deduce that the optimally efficient algorithm has close to zero *Time complexity* and zero *Space complexity*. This paper theorizes that unifying Space and *Time complexity* will give us *Dimensional complexity*. *Time complexity*, *Space complexity,* and *Dimensional complexity* are needed to completely define all finitary algorithms.

With this, we can deduce that an optimally efficient finitary algorithm has zero *Time complexity*, zero *Space complexity*, and an infinite *Dimensional complexity*. Thus finitary algorithms have Dimensional complexities that are less than infinite.

## 3. Completeness and Consistency

Finitary algorithms are useful for solving problems, but there is no fundamental algorithm to compare finitary algorithms to. As aforementioned, This makes calculating the efficiency of a finitary algorithm inconsistent and complicated. A system outlined by Kurt Gödel measures algorithms by their completeness and consistency.

In his paper "On Formally Undecidable Propositions of Principia Mathematica and Related Systems" published in 1931, Kurt Gödel outlines his two theorems of incompleteness and inconsistency. Gödel's theorems outlined the foundational limitations of mathematics and its axiomatic framework [8]. I describe Gödel's theorems as ' bolting a door in front of logic'.

First Incompleteness Theorem:

"Any consistent formal system F within which a certain amount of elementary arithmetic can be carried out is incomplete, i.e., there are statements of the language of F which can neither be proved nor disproved in F".

Second Incompleteness Theorem:

"For any consistent system F within which a certain amount of elementary arithmetic can be carried out, the consistency of F cannot be proved in F itself".

Gödel's first incompleteness theorem stated that no single formal system could completely define all mathematical truths [9]. Gödel's second incompleteness theorem confirmed the unprovability of consistency, by showing that the consistency of mathematics was unprovable within mathematics itself [10]. It is asserting that no mathematical theory can prove its consistency. The result of Gödel's theorems is that mathematics could never completely model language. In 1973 British scientist Alan Turing published a paper "On computable numbers with an application to the Entscheidung problem", where he introduced the Turing machine which formalized an algorithmically computable function [11]. In the paper, Turing used computation to confirm the same relationship demonstrated by Gödel's incompleteness theorems. He demonstrated that the *Entscheidung*, German for "decision problem" was undecidable because no algorithmic process could decide whether an arbitrary mathematical statement was true or false [11]. This problem is known as incomputability and it arises when attempting to design a Turing complete machine also known as a universal computing machine with the power to compute arbitrary algorithms [4]. This proved that it was impossible to algorithmically generate all axioms in a formal system. Essentially a Turing complete machine behaves as an access point to infinite memory or information. Computability theory further outlined that incompleteness is a fundamental property of finitary algorithms. Are infinitary algorithms the most complete and consistent? How do we measure the completeness and the consistency of an algorithm? Essentially, this means that until a complete and consistent theory is deduced to explain the emergence, interactions, and evolution of finitary algorithms that door will remain locked.

If the governing rules of all algorithms are infinitary, then the hypothetical question Gödel will ask is which algorithm is most complete and consistent, finitary or infinitary algorithms? Are finitary algorithms fundamentally incomplete and inconsistent? This paper aims to show that finitary algorithms are fundamentally incomplete and inconsistent without a baseline algorithm as a frame of reference to make them complete and consistent. This baseline algorithm is intelligence, which is informally defined as Infinitesimally-infinitely-finite that is completely-infinitely-incomplete and consistently-infinitely-inconsistent based on the requirements of Gödel's completeness and consistency theorems. This baseline algorithm governs the emergence, interaction, and evolution of all subsequent algorithms.

## 4. Upper and Lower Bounds

In 1892, P. Bachmann's published his book Analytische Zahlentheorie where he introduced the Big-O notation for comparing algorithmic efficiencies by matching an algorithm's growth effort with standard functions [12]. The 5 asymptotic notations are Big-O or *O()*, little-o or *o()*, Big-Omega or *Ω()*, little-omega or *ω()*, and Big-Theta or *Θ().* Essentially *Big-O* or *O()* describes a tight upper bound on the growth of an algorithm and little-o or o() describes an algorithm whose upper bound cannot be tight [12]. Big-Omega or *Ω()* describes an algorithm with a tight lower bound and little-omega or ω() describes an algorithm with a loose lower bound. This paper builds on the work of P. Bachmann and mathematician Donald Knuth's definitions of these notations. In 1976, Donald Knuth published a paper "Big Omicron *O* and big Omega *Ω* and big Theta *Θ* ", where he fine-tuned the definitions of these notations.

When an algorithm has a tight upper bound and tight lower bound, we describe the algorithm with Big-Theta or *Θ()* [12]. Algorithms that admit both Big-O or *O()* and Big-Omega or *Ω()* are said to be asymptotically optimal. These asymptotic notations are represented in **Figure 1** below from CSc 345 - Analysis of Discrete Structures from The University of Arizona. In the figure, n denotes *Space Complexity* and effort denotes *Time complexity*. **Table 1**, shows the limit behavior of corresponding notations. In essence, The notation *f(n) = O(g(n)* where |f| is bounded above tightly by *g*.The notation *f(n) = o(g(n))* where f is bounded loosely above by g. The notation *f(n) = Ω(g(n))* where *f* is bounded tightly below by *g*. The notation *f(n) = ω(g(n))* where f is bounded loosely below by *g*. The notation *f(n) = Θ(g(n))* where f is bounded both tightly above and below by g. The notation *f(n) ~ g(n))* where f is equal to g with a limit of 1 as *lim n-> ∞ f(n)/g(n) = 1.*

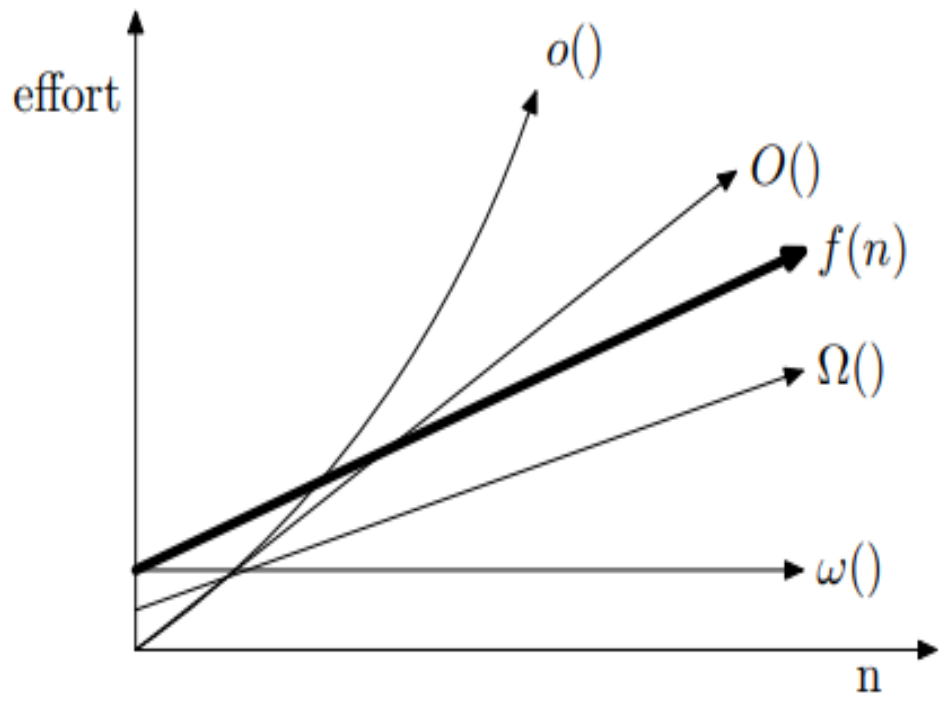

**Figure 1.** Space (*n*) complexity and Time (*effort*) complexity and their relationship with the 5 corresponding asymptotic notations[12].

**Table 1.** Asymptotic analysis of corresponding notations [13].

| Notation | Common name | Limit test |
|---|---|---|
| $f(n) \epsilon O(g(n))$ | Asymptotic Upper Bound | $\lim_{x->\infty} = \left\lvert \frac{f(x)}{g(x)} \right\rvert < \infty$ |
| $f(n) \epsilon o(g(n))$ | Asymptotically Negligable | $\lim_{x->\infty} = \left\lvert \frac{f(x)}{g(x)} \right\rvert = 0$ |
| $f(n) \epsilon \Omega(g(n))$ | Asymptotic Lower Bound | $\lim_{x->\infty} = \left\lvert \frac{f(x)}{g(x)} \right\rvert > 0$ |
| $f(n) \epsilon \omega(g(n))$ | Asymptotically Dominant | $\lim_{x->\infty} = \left\lvert \frac{f(x)}{g(x)} \right\rvert = \infty$ |
| $f(n) \epsilon \Theta(g(n))$ | Asymptotically Tight Bound | $0 < \lim_{x->\infty} = \left\lvert \frac{f(x)}{g(x)} \right\rvert < \infty$ |

We can unify all four notations *o, O, Ω* and *ω,* into two notations that govern a pair of notations each as in notations *Θ* and ~. Where *Θ* is the unified tight bounds as defined by Donald Knuth and ~ is the unified loose bound. When notation *f(n) ~ g(n))* where f is equal to g with a limit of 1 as l*im n-> ∞ f(n)/g(n) = 1,* governing the loose upper bound little-o *o()* and loose lower bound little-omega ω(). The notation *f(n) = Θ(g(n))* where f is bounded both tightly above *O()* and tight below *Ω()* by g. Mathematician Donald Knuth's 1976 paper showed that the limit behavior of *f(n) = Θ(g(n))* is both *f(n) = O(g(n))* and *f(n) = Ω(g(n))* [14].

Knuth defined as:

$$f(x) = \Omega(g(x)) \Leftrightarrow g(x) = O(f(x)) \tag{1}$$

This means the limit *lim n-> ∞ f(n)/g(n)* of notation *f(n) = Θ(g(n))* is greater than zero but less than infinity (0, ∞). Essentially *f(n) ~ g(n)* is infinitary or continuous and *f(n) = Θ(g(n))* finitary or discrete. Unifying finitary and infinitary algorithms reduces to the equal sign notation " = ". Knuth's established the equivalence between the 5 notations *o, O, Θ, Ω, ω,* and the number line [15]. Knuth's unification notation *(Θ)* is equivalent to the " ≈ " on the real number line as shown in **Figure 2** below. Despite the fact that " = " and the " ≈ " fit on the real number line " <, <=, ≈, =, >=, > ", We aim to simplify the notations further. The " = " governs the infinitary algorithm and the " ≈ " governs the finitary algorithm.

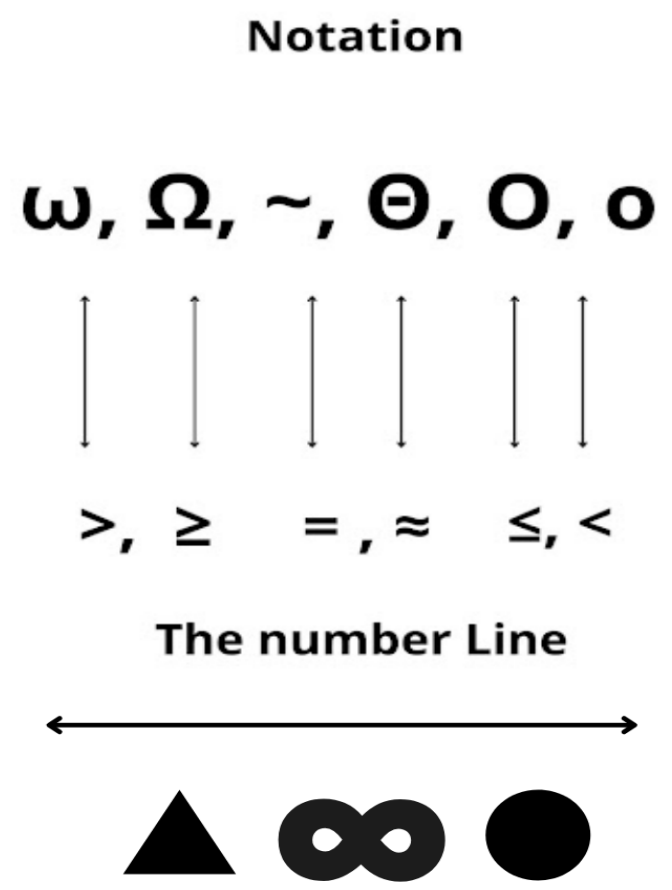

**Figure 2.** The number line with corresponding notations and formalized abstraction of intelligence.

Unifying finitary *(Θ)* with infinitary (~) is equivalent to unifying the notations " = and ≈ ". We aim to simplify the six notations below into three notations while preserving the number line as seen in my illustration in **Figure 2** below. The notation that I present in this paper is **Δ ∞ O**, where the notation O denotes finitary algorithms, the notation Δ denotes infinitary algorithms and the notation ∞ denotes infinite dimensionality. The 5 aforementioned notations *o, O, Θ, Ω, ω,* the number line, and the formal definition of intelligence are used to define a complete and consistent Intelligent Engine or Turing complete machine as shown in the illustrations I designed in **Figure 3**-7. Thus intelligence is an abstraction or formalization of generality.

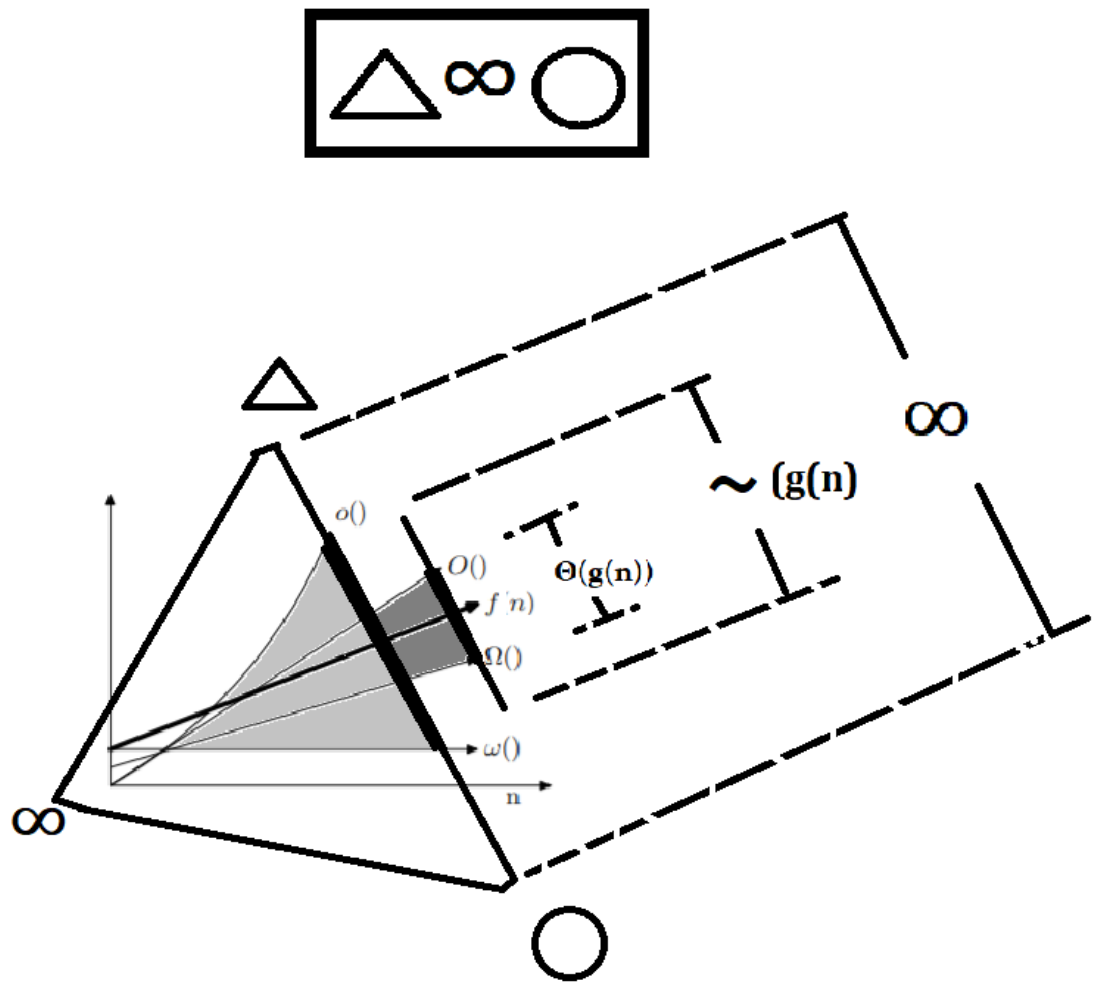

**Figure 3.** Algorithmic Intelligence is defined formally as **Δ∞ O** and informally as Infinitesimally-infinitely-finite. (Δ denotes *Time complexity* ∞ denotes *Dimensional complexity* and O denotes *Space complexity*).

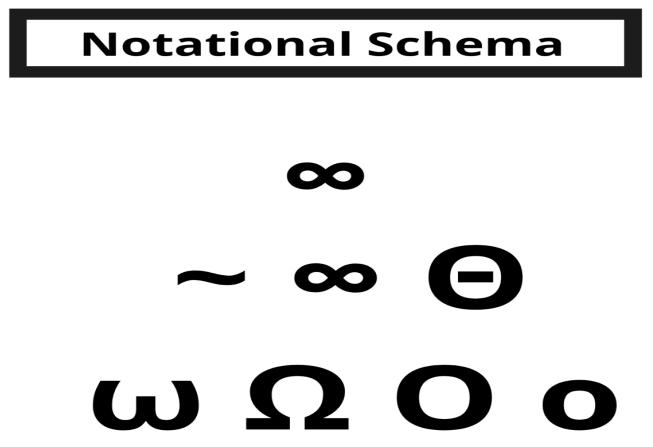

**Figure 4.** Notational Pyramid of Intelligence.

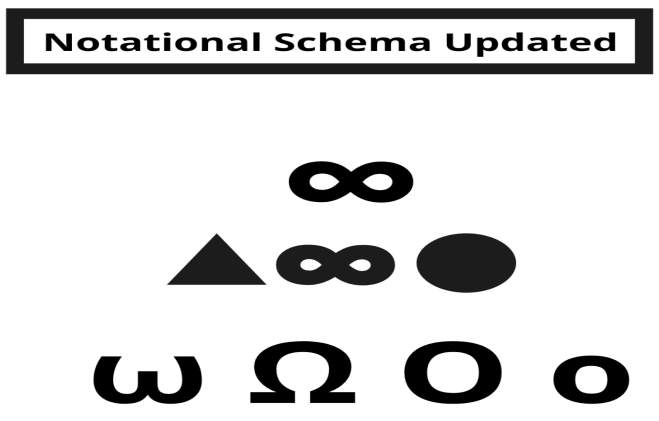

**Figure 5.** Updated Notational Pyramid of Intelligence.

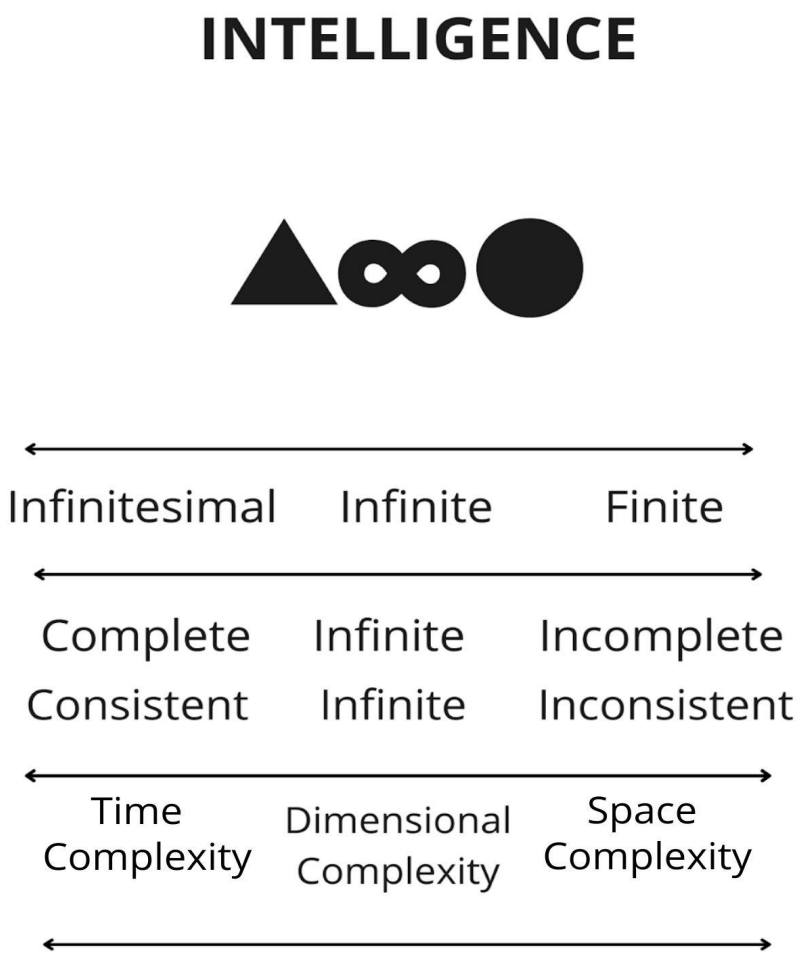

**Figure 6.** Intelligence. Defined Formally, Informally, and Computationally
( Δ connotes infinitesimal and denotes *Time complexity*. ∞ connotes infinity and denotes *Dimensional complexity*. O connotes finite and denotes *Space complexity*).

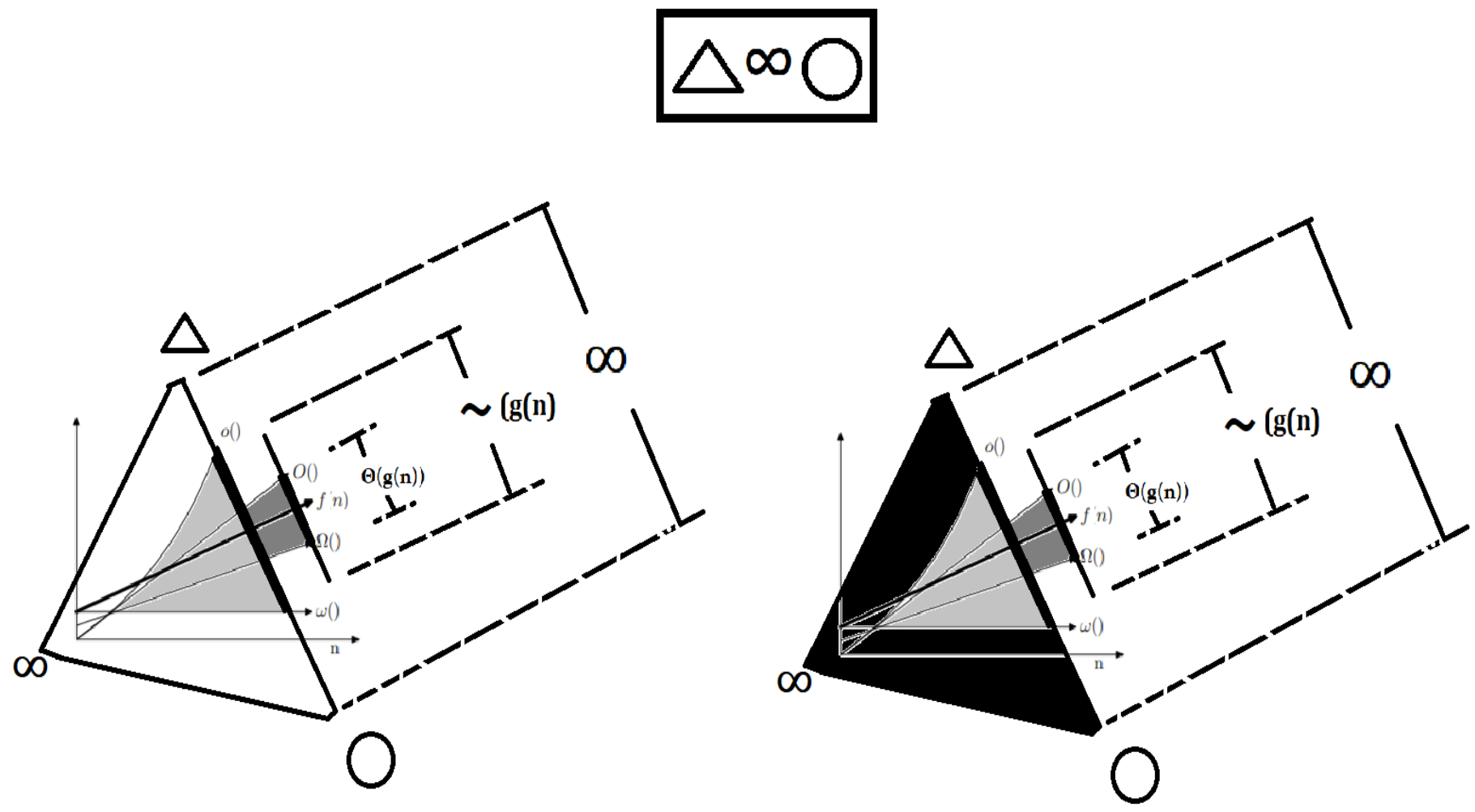

**Figure 7.** Intelligence Engine. Defined formally as **Δ∞O** and Informally as *Infinitesimally-infinitely-finite*.

## 5. From the “ = ” to the “ ∞ ”

Robert Recorde, circa 1510 to 1558, is usually cited as the first to use the equal-to-sign “ = ” symbol in his work. Recorde used the two parallel lines to represent equality between two things [16]. Before the equal sign came into common use, there were other forms of expression of equality. In Florian Cajori's work A History of Mathematical Notations: Vol.1—Notations in Elementary Mathematics, pages 297–298, he showed that the “ = ” sign was not generally accepted in academia until 1631. It was adopted as the symbol of equality in some influential works in England including Thomas Harriot’s Artis analyticae praxis, Willian Outhtred’s Clavis Mathematicae, and Richard Norwood’s Trigonometria [17].

Symbols are used as a shorthanded way of simplifying ideas. They are discrete representations of continuous ideas. The equal Sign “ = ” is an abbreviation of expressing the idea “ is equal to” symbolically. In mathematics, “ : = “ means "is denoted by" or "is defined as". Asserting equality “ = ” does not mean the same thing as denoting “: =”, which does not assert equality. For example, X denotes Y is represented formally as X: = Y, and X is equal to Y is represented formally as X = Y. The ambiguity between X : = Y and X = Y has created a pervasive use of the equal sign “ = ” in symbology. Essentially the symbol of equality “ = ” has been incorrectly used to mean “denote” and this misuse has effectively complicated our understanding of the relationship between finitary and infinitary algorithms. Unifying finitary and infinitary algorithms comes down to the definition of the equal sign notation “ = ”.

In Steven G. Krantz’s 2016 paper titled A Primer of Mathematical Writing, he writes ” The dictionary teaches us that “A connotes B” means that A suggests B, but not in a logically direct fashion” [18]. Denotation implies an explicit relationship and connotation implies an indirect relationship. Denotation and connotation co-exist but denotation is primary and connotation is secondary. This paper builds on the definitions of Recorde and Knuth by unifying connotation ” = : ” and denotation “ : = “. In summary, unifying both will mean finding the relationship between “ : = “ and “ = : ”, where connotation ” = :” or “ = ” is infinitesimal and denotation “ : = “ or “≈” is finite as shown in **Figure 2**. The Informal definition of intelligence as aforementioned is “infinitesimally-infinitely-finite” and the formal definition is “ **Δ ∞ O** ”.

## 6. Dimensional Complexity

This paper claims that a complete and consistent model of algorithmic efficiency requires *Time Complexity*, *Space complexity,* and *Dimensional complexity*. We now use our algorithm of intelligence informally defined as infinitesimally-infinitely-finite and formally defined as **Δ ∞ O** to formalize *Time complexity*, *Space complexity,* and *Dimensional complexity*. *Space complexity* O has finite dimensionality, *Time complexity* Δ has infinitesimal dimensionality, and *Dimensional complexity* ∞ has infinite dimensionality.

It is deduced that this optimally efficient baseline algorithm has zero *Time Complexity*, zero *Space complexity*, and an infinite *Dimensional complexity* **Δ ∞ O**. This generalized abstraction of intelligence is illustrated in **Figure 6** and **Figure 7**, where I outline the Triarchic relationship between **Δ ∞ O**.

In summary:
**Δ** connotes infinitesimal and denotes *Time complexity*.
**∞** connotes infinite and denotes *Dimensional complexity*.
**O** connotes finite and denotes *Space complexity*.

## 7. Conclusions

The purpose of this paper was to outline a theoretical framework of intelligence, informally, formally, and computationally. The concept of *Dimensional complexity* in algorithmic efficiency was introduced and it was deduced that an optimally efficient algorithm has zero *Time complexity,* zero *Space complexity*, and infinite *Dimensional complexity*. In essence, This paper shows intelligence is the algorithm of abstraction or a formalization of generality. The Algorithm is **Δ∞O**, where Δ connotes infinitesimal and denotes *Time complexity*. ∞ connotes infinite and denotes *Dimensional complexity.* O connotes finite and denotes *Space complexity.* This algorithm was used to recreate the number line.

This paper shows that finitary algorithms have Dimensional complexities that are less than infinite which limits Time and *Space complexity* to non-zero. Hence the reason finitary algorithms are essentially birthed by infinitary algorithms. In computation, *Dimensional complexity* is the non-infinite data set or input size of information that fundamentally limits the efficiency of an algorithm by increasing time and *Space complexity*. Future works include applying the generalized principles of intelligence from this paper in system design and optimization problems. Based on the issues discussed here, it is proposed that mathematics, computer science journals, and institutions adopt the **Δ∞O** relations of intelligence as defined above.

## 8. Implication: The Genesis Algorithm

This study implies that a generalized symbolic abstraction of intelligence "**Δ∞O**" is complete and consistent. This fundamental representation of intelligence acts as an access point or conduit to infinite intelligence. Future research includes using this algorithm to access this intelligence.

## Conflicts of Interest

The author declares no conflicts of interest regarding the publication of this paper.

## References

[1] Hosch, W.L. (2022) Incompleteness Theorem. Encyclopedia Britannica.

[2] Panu, R. (2013) Gödel's Incompleteness Theorems. Stanford Encyclopedia of Philosophy.

[3] Church, A. (1936) An Unsolvable Problem of Elementary Number Theory. American Journal of Mathematics, **58**(2), 345–363. https://doi.org/10.2307/2371045

[4] Turing, A.M. (1937) On Computable Numbers, with an Application to the Entscheidungsproblem. *Proceedings of the London Mathematical Society*, **42**, 230-265. https://doi.org/10.1112/plms/s2-42.1.230

[5] Kushner, B. A. (2006). The Constructive Mathematics of A. A. Markov. The American Mathematical Monthly, **113**, 559–566. https://doi.org/10.2307/27641983

[6] Markov, A.A. 1960. The Theory of Algorithms. American Mathematical Society Translations, series 2, **15**, 1-14. http://doi.org/10.1090/trans2/015/01

[7] Kuo, W. and Zuo, M.J. (2003) Optimal Reliability Modeling: Principles and Applications. John Wiley & Sons, Hoboken, 62.

[8] Juliette, K. (2020) Kurt Gödel. Stanford Encyclopedia of Philosophy, Plato.stanford.edu. https://plato.stanford.edu/entries/goedel

[9] Davis, M. (2006) The Incompleteness Theorem. *Notices of the AMS*, **53**, 414..

[10] Panu, R. (2020) Gödel's Incompleteness Theorems. Stanford Encyclopedia of Philosophy.

[11] Cook, S.A. (2007) An Overview of Computational Complexity. ACM Turing Award Lectures. Association for Computing Machinery, New York .https://doi.org/10.1145/1283920.1283938

[12] McCann, L. (2009) Analysis of Discrete Structures. University of Arizona CSc 345, 1-2.

[13] Ellefsen, B. (2020) Calculating big-$O(x3)$. https://math.stackexchange.com/q/2126104

[14] Knuth, D.E. (1976) Big Omicron and Big Omega and Big Theta. *SIGACT News*, **8**, 18-24. https://doi.org/10.1145/1008328.1008329

[15] Vitányi, P. and Meertens, L. (1985) Big Omega versus the Wild Functions. *ACM SIGACT News*, **16**, 56-59. https://doi.org/10.1145/382242.382835. S2CID 11700420

[16] Seehorn, A. (2021) The History of Equality Symbols in Math. sciencing.com.

[17] Cajori, F. (1928) A History of Mathematical Notations: Vol.1—Notations in Elementary Mathematics. The Open Court Company, London, 297-298.

[18] Krantz, S. G. (2016). A primer of mathematical writing.

Dedicated to the Kosso and Ngu family
of Cameroon, Africa.